# Magnesium hydride films deposited on flexible substrates: structure, morphology and hydrogen sorption properties

Huy Le-Quoc [1,2+], Marie Coste [1,2], Ana Lacoste [2], Laetitia Laversenne[1*]

[1] Univ. Grenoble Alpes, CNRS, Grenoble INP, Institut Néel, 38000 Grenoble, France

[2] LPSC, Université Grenoble-Alpes, CNRS/IN2P3, 53 rue des Martyrs, 38026 Grenoble, France

[+] Present address: University of Danang, University of Science and Technology, Danang 550000, Vietnam

* Corresponding author

**Abstract**

1.8 micrometer-thick magnesium hydride films were synthesized in a single-step process by reactive plasma-assisted sputtering. The $MgH_2$ thin films, which were deposited on two types of flexible surfaces (namely graphite and polyimide foils) were found to adhere on both substrates. In all cases, XRD analysis revealed an as-deposited thin film consisting of *alpha*-$MgH_2$, a tetragonal, rutile-type crystal structure (space group #136). The hydrogen sorption capacities of the uncapped films were studied over successive desorption/absorption cycles performed at 350°C. The first desorption always shows a slow kinetics that can be explained by a superficial oxidation of the films. However, once the passivating layer is removed, the following dehydrogenations occur faster. Multiple cycling of the film deposited on polyimide resulted in delamination of the film and its conversion into loose powder. As for $MgH_2$ deposited on the flexible graphite substrate, a fully reversible capacity was observed over 28 cycles with no delamination of the film. Upon cycling, the microstructure of the film has evolved from homogeneous fibrous to an untextured morphology with a higher degree of crystallinity.

**Corresponding author**:

L. Laversenne, laetitia.laversenne@neel.cnrs.fr

# 1. Introduction

Ultrathin devices based on flexible substrates are receiving growing attention as they offer solutions to meet new technological demands. For instance, flexible electronics are suitable for portable and wearable equipment as they are lightweight, they minimize dead volumes, and their conformable properties allow them to be transferred onto any surface or object. Another major asset of ultrathin devices is their cost-effectiveness which is due to the minimized amount of consumed raw materials and to the fact that they can be manufactured by high-productivity roll-to-roll technologies. While functional films composed of organic materials deposited on plastic films have been implemented for many years, the development of inorganic materials on flexible substrates is much more recent. Among the latter, one can cite piezoelectrics [1], supercapacitor electrodes [2], thermoelectrics [3], and anodes for rechargeable batteries [4]. As an example of metal hydrides deposited on flexible substrates, LaNi$_5$ alloys were proposed for applications as actuators and sensors which use the large volume change upon hydrogen absorption [5]. Additional potential applications of metal hydrides rely on the significant changes in the electronic and optical properties which are induced upon the reversible transition between the (insulating) hydride and the metal. Among investigated systems, yttrium, lanthanum and magnesium-based films deposited on rigid substrates have been considered for switchable mirrors, sensors and smart windows [6–12]. In the past, we provided a proof of concept of direct growth of MgH$_2$ films by using Plasma-Assisted Physical Vapor Deposition (PAPVD) under reactive conditions [13]. The approach was successfully applied to deposit MgH$_2$ over Cu current collector in the attempt to develop high-capacity anode materials for Li-ion batteries – a study in which the late Michel Latroche had taken part [14]. Magnesium and magnesium-based compounds are benchmark, *i.e* reference, materials for hydrogen storage [15–21]. Yet, as for MgH$_2$ obtained as films, they have received much less attention than powders. In fact, these materials with such a specific shaping have been mainly considered to unravel theoretically [22–24] and experimentally [25–31] fundamental aspects of hydrogen sorption properties. In this paper, we report on the properties of MgH$_2$ micron-sized films deposited onto flexible substrates, namely polyimide foil and graphite sheets. The two substrates were selected for their chemical inertness when set in contact with Mg and MgH$_2$. They are also cheap and readily available, which is an asset for possible applications. In addition to showing the possibility to deposit MgH$_2$ films on such flexible substrates, we also report on the structural and microstructural properties of the films as well as their hydrogen desorption-absorption cycling capabilities.

## 2. Experimental Methods

*2.1. Deposition of magnesium hydride films*

400 µm-thick graphite foil (Sigraflex® from SGL Group, density = 0.75 g/cm$^3$) and 12.5 µm-thick polyimide foil (HN 50 type Kapton®, density = 1.42 g/cm$^3$) were used as flexible substrates. Prior to use, the surfaces of the polyimide foil were cleaned in an ultrasound bath with acetone and ethanol for 5 min each. The graphite sheet was heat-treated at 800°C under vacuum for 2 hours. The close contact between the substrate and the holder was ensured by a mechanical assembly which keeps the two elements pressed together with high thermal transfer. We noticed that the contact interface is a particularly critical point in the case of thermally insulating substrates such as polyimide. In the deposition chamber, water cooling of the substrate-holder was used to prevent heating of the sample during film growth. Before deposition, the surfaces of the substrate and targets were cleaned successively using an Ar plasma (at a pressure of 0.2 Pa) with a -55 V RF voltage (for the substrate) and a -200 V DC voltage (for the target). Magnesium hydride films were deposited by reactive plasma sputtering according to the pulsed procedure described in [13] at a vacuum base pressure of 10$^{-5}$ Pa. An Ar-H$_2$ mixture with partial pressure of 0.2 Pa Ar and 0.2 Pa $H_2$ was introduced in the vacuum chamber. Independent bias of the substrate and targets ensures an accurate control of both the flux and energy of deposited particles. While the substrate was electrically floating, the three Mg (99.99%) targets were polarized by a -200 V DC voltage. Deposition times were varied between 2 and 3h leading to magnesium hydride film with a thickness of between 1 and 2 µm.

*2.2. Thin film characterization*

Apart from the unavoidable exposure to air when the sample is taken out from the deposition chamber, the films were handled and stored exclusively under inert atmosphere in a glove box. Their microstructure and thickness were investigated using a field emission scanning electron microscope (ZEISS-Ultra +). Phase identification was performed by means of transmission X-ray diffraction (XRD) with a Bruker D5000T diffractometer (for films deposited on polyimide) and in-grazing incidence geometry using a Bruker D8 Advance diffractometer (for films deposited onto graphite). In both cases, diffractograms were collected at the wavelength λ = 1.54 Å with a $\theta$ step of 0.032° and time step of 20 s. The hydrogen sorption properties of the films were studied using a HERA volumetric (Sievert's type) apparatus. The non-reactivity of the bare substrates was checked under the same conditions as the sorption measurements. The exact volume of film introduced in the sample holder was calculated by multiplying the film thickness and the precisely determined sample surface (about 50 $cm^2$ in total). Hydrogen desorption and absorption measurements were performed at a constant temperature of 350°C under an initial hydrogen pressure of 4 kPa and 1040 kPa, respectively. The duration of the first three desorption/absorption steps was set manually, while the subsequent absorptions were automatically stopped at 86% of the storage capacity to save time. The quantitative determination of the hydrogen storage capacity of the films was possible with a fine calibration of the Sievert's apparatus. The configuration of our experiment allows us to obtain an accuracy of the measurement of 0.2 kPa which corresponds to a hydrogen amount of $n = 8.438$ µmol $H_2$. We point out that sorption kinetics can only be considered qualitatively from the present data because the sample (rectangular pieces of film + substrate) occupies a large volume of the cylindrical sample holder but has a limited number of contacts with the thermalized wall. For this reason, heat transfers that are associated with $H_2$ sorption reactions are slowed down. It is known that thorough thermal management is required to deal with the enthalpy of formation of $MgH_2$ and optimize reaction kinetics [17].

## 3. Experimental results

The $MgH_2$ films were successfully deposited on the two types of substrates, namely polyimide and graphite foils. In all cases, the as-deposited films show uniformity and good adhesion to the substrate over the whole surface. For illustrative purposes, two samples are shown in Figure 1. As can be seen in the figure, the polyimide sample tends to roll up when removed from the substrate holder after deposition, therefore indicating that the $MgH_2$ film is on tensile stress. In contrast, such rolling up is not observed with films deposited on the graphite substrate (this point will be commented in more detail later in the discussion section). Measurement of the film thickness along the diameter shows a continuous reduction that reaches 15% at the edge. The thickness profile, which is directly related to the geometric configuration of the deposition chamber, can be described by a pseudo-Voigt function used to calculate the volume of the film characterized by sorption measurements. In what follows, we first discuss the data obtained for the polyimide substrate and then for the graphite substrate.

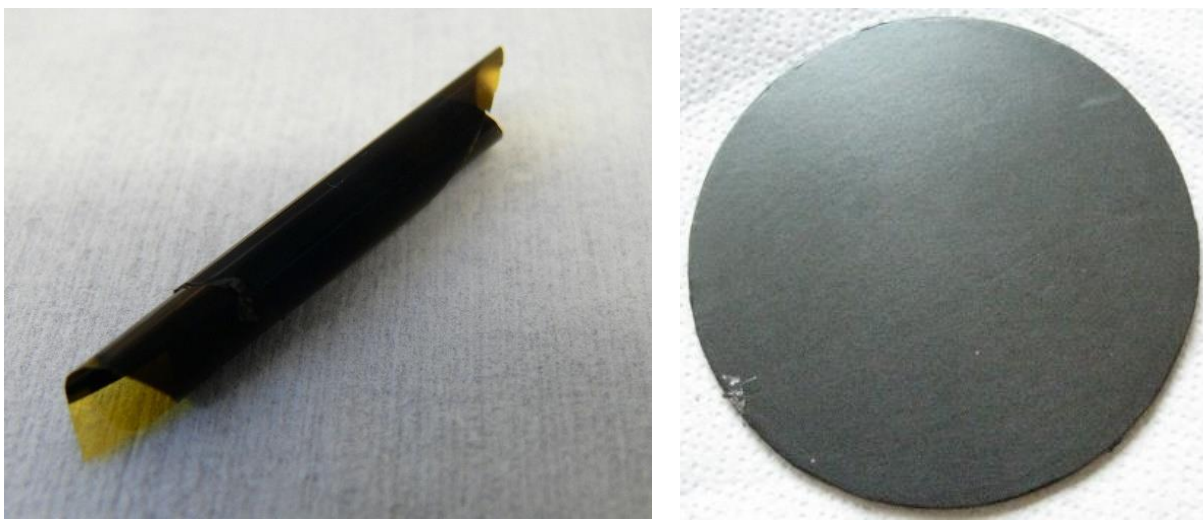

*Figure 1* – *Magnesium hydride films deposited on a polyimide foil (left) and on a 10 cm diameter flexible graphite disk (right).*

**MgH$_2$ films deposited on a polyimide foil.** A strong adhesion of the as-deposited MgH$_2$ films on the polyimide foil was observed. Whereas polymer metallization usually requires a preliminary adhesion promoting process [32], the only preparation treatment in the present work was the "*in situ*" Ar plasma cleaning step applied for 10 minutes (as described in the experimental part). The strength of the adhesion was qualitatively assessed by manual scratching with tweezers and attachment/removal of adhesive tape. No delamination was observed, therefore demonstrating the strong interaction between the polyimide and the as-deposited film. SEM observation of a tilted sample reveals that the as-deposited MgH$_2$ film has a uniform morphology over the entire thickness (cf. Fig 2, *left*). The top surface of a second film can be visualized in the high magnification (×50 k) micrography in Fig 2, *right*. The average grain intercept method [33], which was applied to the latter image, leads to a mean grain size value of 117(15) nm. The largest and smallest grains visible on the top surface are respectively 345 and 40 nm-long.

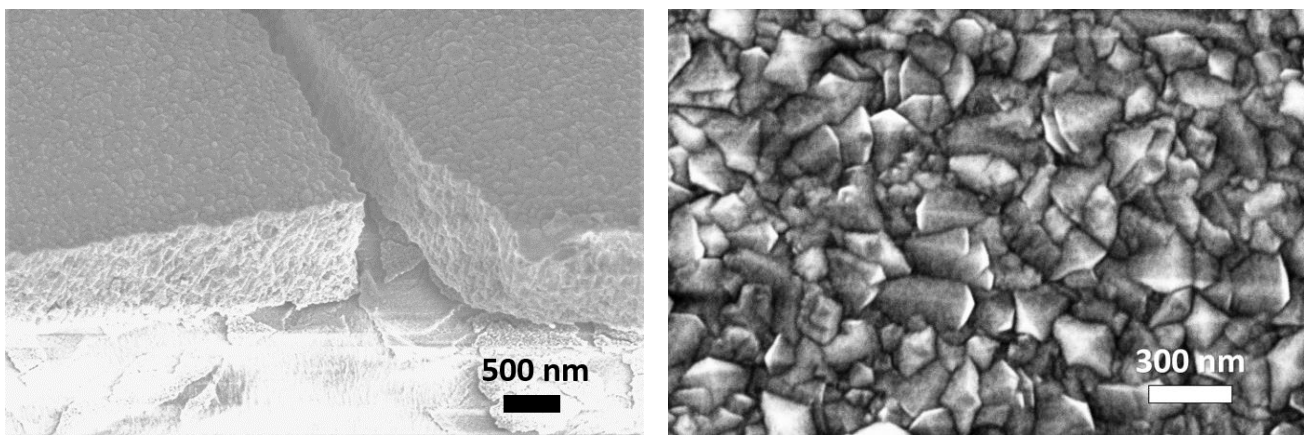

*Figure 2* – SEM micrographs of magnesium hydride films deposited on a polyimide substrate. Tilted image taken at ×20 k magnification showing both the cross section and top surface (left). The SEM micrograph on the right, which is obtained for the sample investigated in this study, shows the top surface image collected at a ×50 k magnification.

The XRD pattern collected for the $MgH_2$ film deposited on a 12.5 μm-thick polyimide foil is shown in **Figure 3**. We also show the XRD pattern obtained for the bare substrate. Both diffractograms show a high background at low angle due to air scattering. Even with a low signal to noise ratio (due to the small sample volume), the film pattern can be indexed with the Bragg reflections of *alpha*- $MgH_2$ (space group # 136; $a$ = 4.517(2) Å; $c$ = 3.020(2) Å). No traces of crystalline MgO is visible in the XRD pattern. The peak at $2\theta$ = 38.1° ($d$ = 2.36 Å) is thought to be due to a measurement artefact – in any case, it cannot be assigned to a Mg-containing phase (e.g. Mg, *gamma*-$MgH_2$ (space group #60), MgO). The relatively broad width of the observed Bragg peaks is consistent with the nanometer size of the crystallites which has been estimated to be in the range of 20 nm by applying Scherrer's equation."

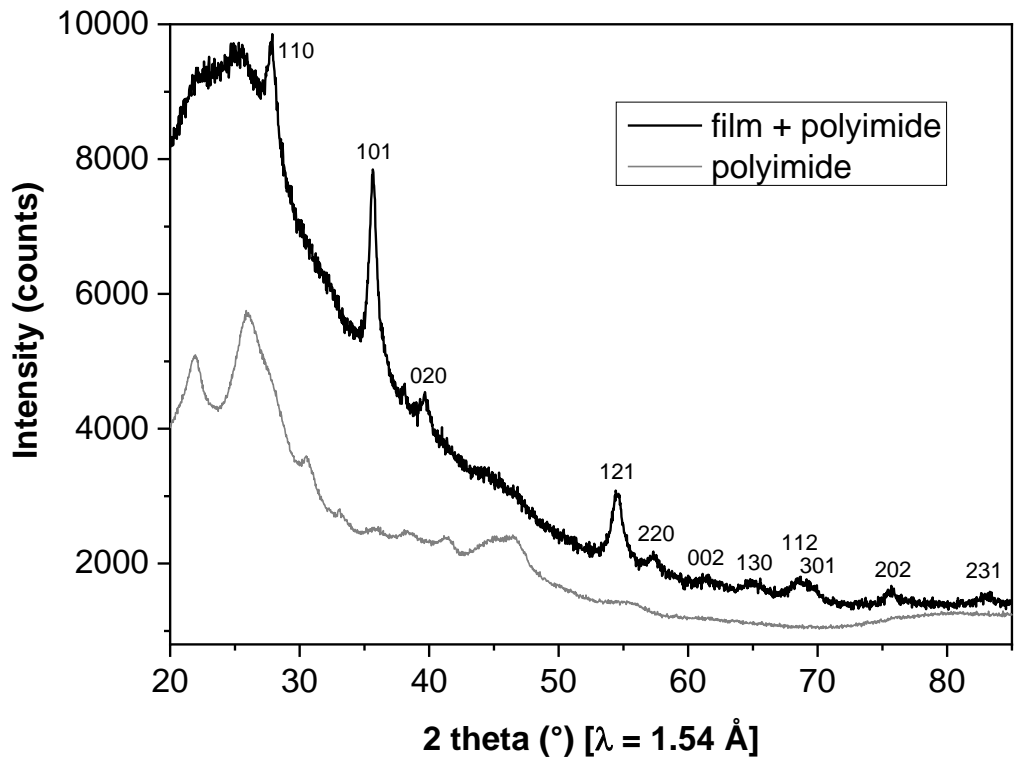

*Figure 3* – *XRD diagram of the magnesium hydride film deposited on a polyimide substrate.*

The hydrogen sorption properties of a 1.78 μm-thick film were investigated by volumetric measurements. Three successive steps (desorption–absorption–desorption) resulted in a film still bonded to the polyimide sheet. However, the film-substrate adhesion (tested using adhesive tape) was weaker than for the film as-deposited. Further cycling of the film over 43 successive desorption–absorption processes (Figure 4) showed no loss in the hydrogen capacity, therefore confirming full reversibility of hydrogen storage and the storing material. The hydrogen gravimetric capacity of the 1.78 μm-thick magnesium film was determined to be 7.4 ± 0.1 wt.%. This value is very close to the full capacity of magnesium (7.6 wt. %). Taking into account the 12.5 μm-thick polyimide substrate, the storage capacity of the sample (film + polyimide) is reduced to 0.97 wt.% $H_2$. The opening of the sample holder after cycling revealed the delamination of the film and its pulverization.

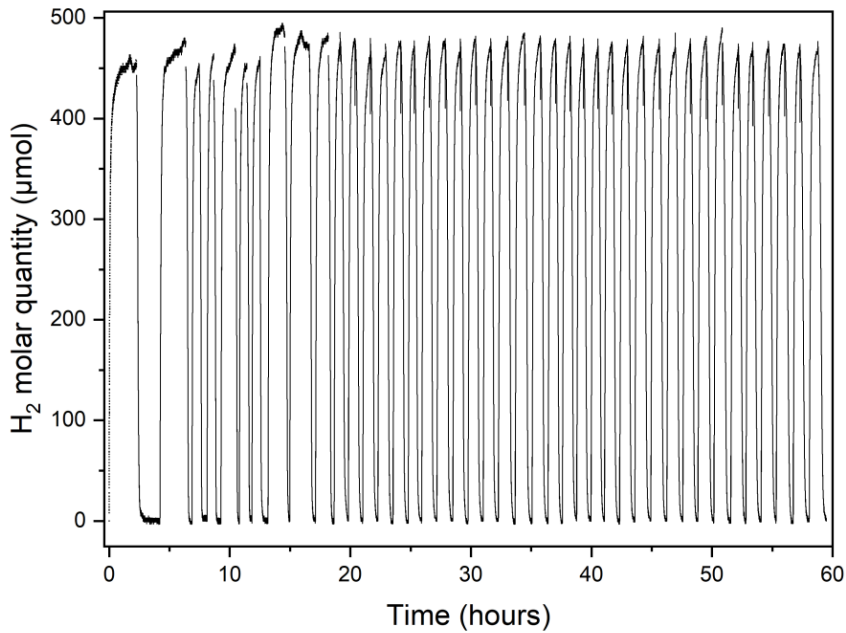

*Figure 4* – *Successive hydrogen desorption/absorption sequences recorded at 350°C (4 kPa and 1040 kPa starting $H_2$ pressure, respectively) for a magnesium hydride film deposited on polyimide.*

**$MgH_2$ films deposited on a graphite sheet.** The morphology of the $MgH_2$ films deposited on graphite sheet was investigated by SEM. The top surface of the films (not shown here) is rather uneven as it follows (*i.e.* reproduces) the roughness of the graphite surface itself. The cross section of the as-deposited 1.8 µm-thick magnesium hydride film presents a compact microstructure made of fibrous grains elongated in the direction perpendicular to the substrate. The width of these features is of the order of 100 nm as can be seen from Figure 5. The fibrous morphology is accompanied by a preferential grain orientation with the (101)-type crystal planes oriented parallel to the sample surface. This can be seen by comparing the relative intensity of (101) film diffraction peak with that of the simulated model for unoriented $MgH_2$ powders (cf. Fig. 6). Lebail profile matching of the XRD pattern allows determining the lattice parameters of *alpha*- $MgH_2$ (space group # 136; *a* = 4.517(1) Å; *c* = 3.017(2) Å). Moreover, the use of Scherrer's equation leads to an estimation of the crystallite size of 22(3) nm, which is comparable to the results obtained for the polyimide-deposited film using the same method.

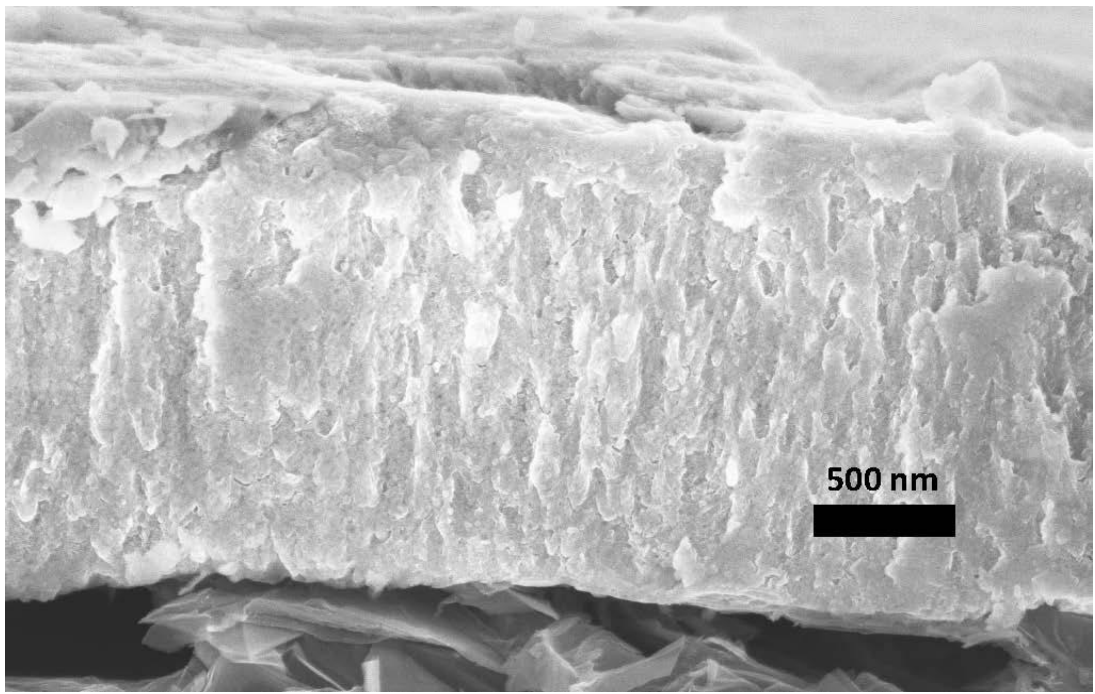

*Figure 5* – *Cross sectional SEM images of the as-deposited magnesium hydride film supported on graphite.*

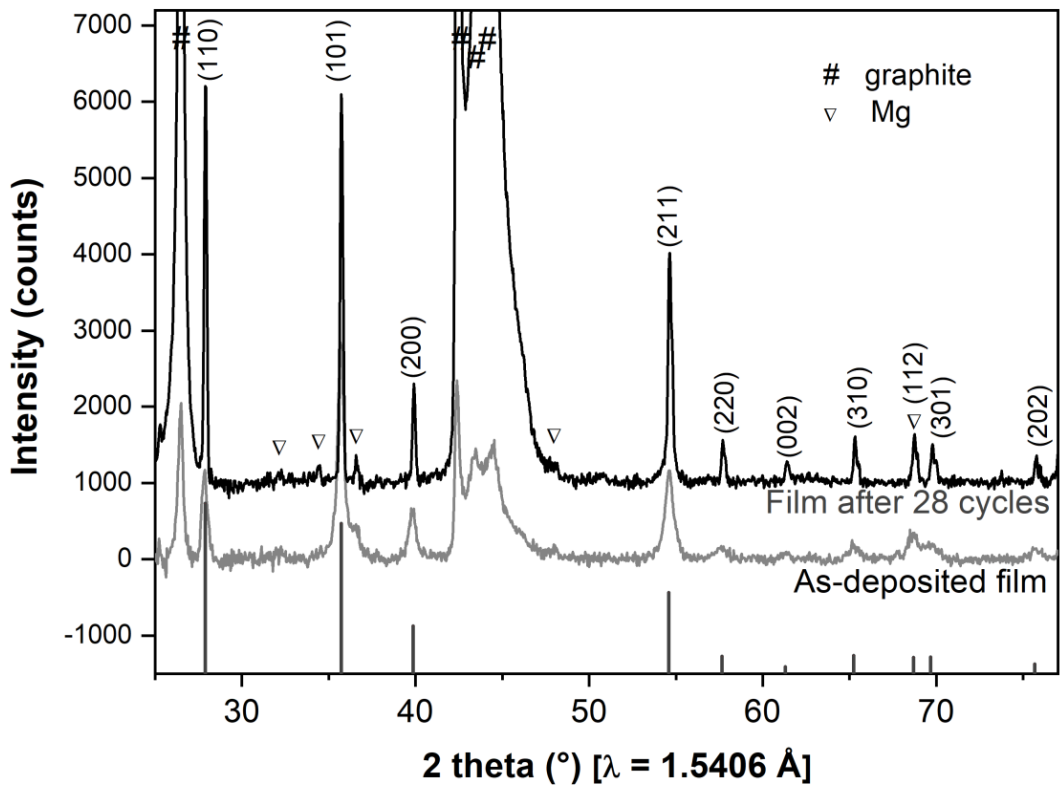

*Figure 6* – *XRD patterns collected on a graphite supported magnesium hydride film as-deposited and after 28 desorption/absorption cycles. Indexed peaks are assigned to alpha-MgH$_2$ whereas triangle and hash symbols stand for Mg and graphite. The relative intensity of MgH$_2$ diffraction peak in randomly oriented grain are displayed in the bottom.*

The results of volumetric measurements for the three first hydrogen desorption-absorption sequences are displayed in Figure 7. The left axis shows the differential pressure between the sample chamber and the reference chamber while the right axis indicates the corresponding hydrogen amount. As can be seen, the first desorption occurs relatively slowly (completed in 2 hours) due to the presence of a passivation layer which was formed by oxidation at the exit of the pulverization chamber (the film has not been decorated with a protective coating). Once the passivation layer is removed, solid-gas reaction is no longer hindered and subsequent desorptions occur within 20 minutes. The sample was then cycled for an additional 25 times to obtain the molar amounts of hydrogen displayed in Figure 8. The hydrogen gravimetric content of the hydrided magnesium film alone was estimated as 7.4 ± 0.1 wt.%. Considering the supporting graphite substrate, the capacity of the whole system (film + substrate) is 0.07 wt.% $H_2$. Note that the ninth cycle (plotted in gray) has a reduced amplitude due to an unintentional stop before the end of the desorption. As can be seen from the plot, the $MgH_2$ film shows quasi-full cyclability within the multiple cycles. After the 28 sorption cycles, the film is still supported on the graphite substrate in contrast to what was obtained for the polyimide substrate (see results above).

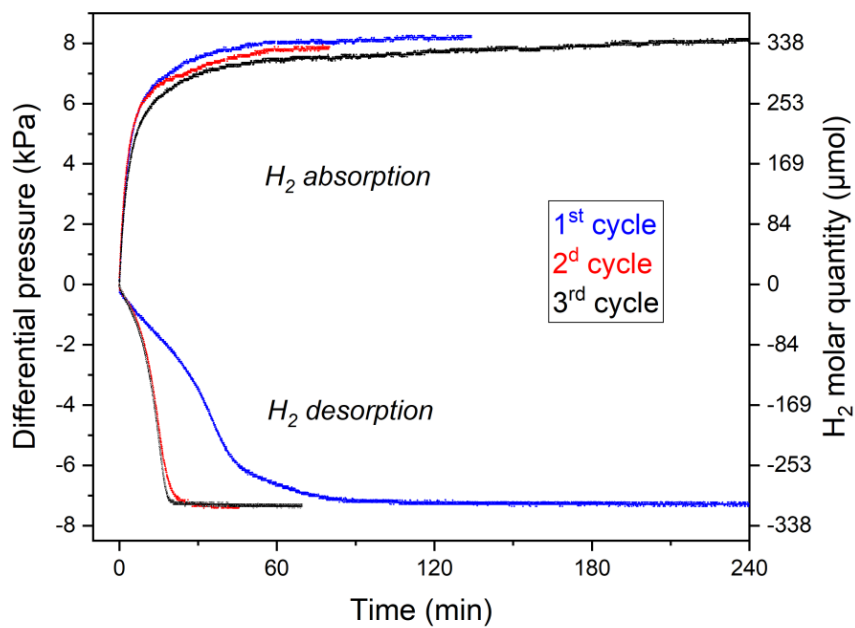

*Figure 7* – *First three desorption/absorption curves recorded at 350 °C (4 kPa and 1040 kPa starting $H_2$ pressure respectively) for a magnesium hydride film deposited onto graphite.*

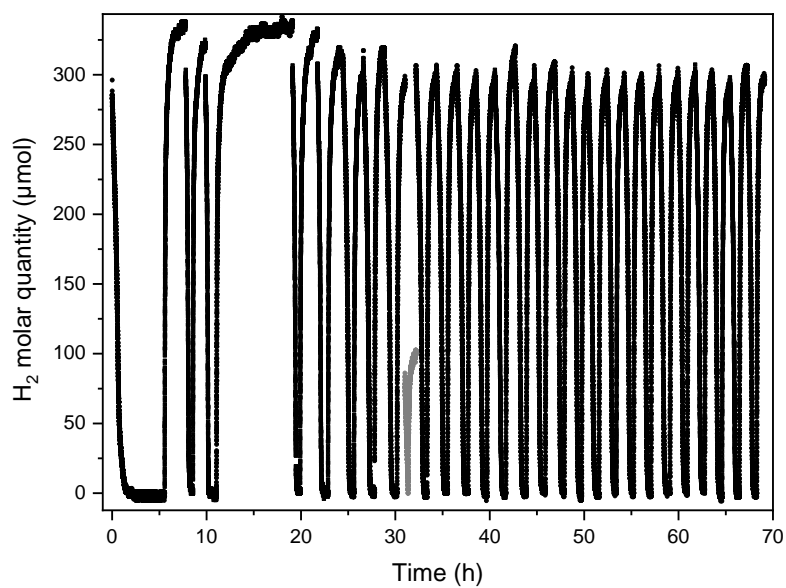

*Figure 8* – *Successive hydrogen desorption/absorption sequences recorded at 350 °C (4 kPa and 1040 kPa starting $H_2$ pressure respectively) for a magnesium hydride film deposited onto graphite. Data in gray color correspond to a measurement for which the desorption process was incidentally stopped.*

SEM observation on the recovered cycled sample is shown in Figure 9. The smooth structures visible at the surface result from sublimation/condensation of magnesium during the cycling process. A similar surface modification was reported for vanadium-activated magnesium hydride powder cycled 2000 times [34]. The SEM image reveals changes in the microstructure – notably with a loss of the fibrous morphology in favor of an apparently denser core structure of the film. This morphology, which is already visible after the first hydrogen sorption cycle, is attributed to the successive crystallizations of the metal and hydride phases. The XRD pattern collected in Figure 6 for the film after 28 cycles (1) reveals a loss of preferential orientation and (2) confirms the film rehydrogenation. No significant changes in the lattice parameters of $MgH_2$ are observed (space group # 136; $a = 4.516(1)$ Å; $c = 3.015(1)$ Å). On the other hand, a small amount of crystallized magnesium is visible in the pattern. The hydrogen sorption cycling is found to be responsible for an increased degree of crystallinity as demonstrated by the enhanced diffraction signal and Bragg peaks sharpening. The mean crystallite size as estimated from Scherrer's equation is increased from 22(3) nm to 54(4) nm after 28 hydrogen desorption/absorption cycles.

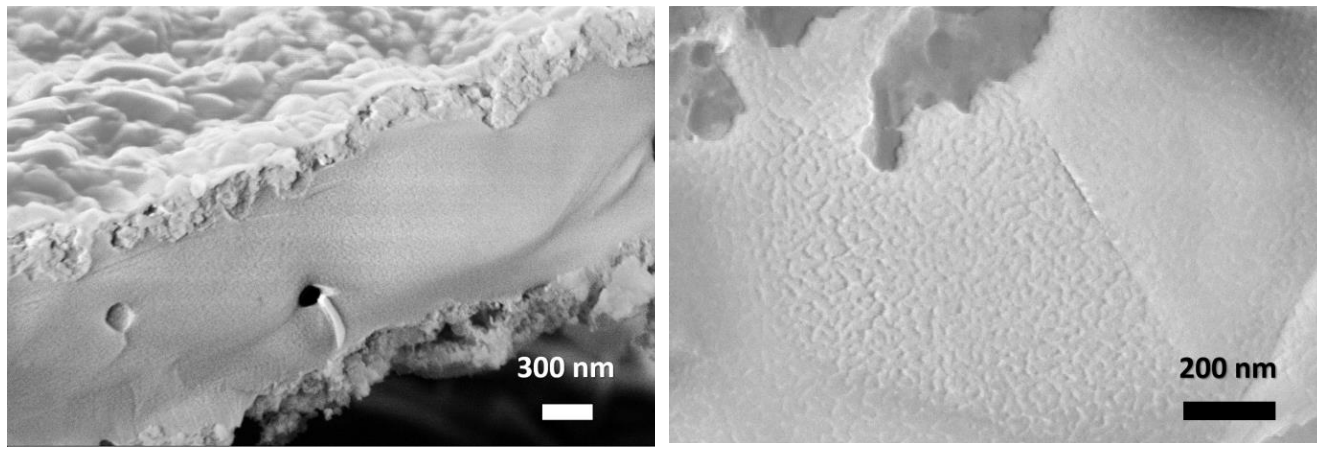

*Figure 9* – SEM cross-sectional images of the cycled magnesium hydride film supported on graphite for two magnifications: ×30k (left) and ×80k (right).

## 4. Discussion

The MgH$_2$ films deposited on graphite and polyimide show a similar degree of crystallinity but a preferential growth orientation is observed in the case of the graphite-supported film. We previously reported a strong texture of Si-supported MgH$_2$ [13]. The order of magnitude difference in thermal conductivity of the substrate ($\lambda \sim 0.2$ W. m$^{-1}$. K$^{-1}$ for polyimide; $\lambda \sim 7$ W. m$^{-1}$. K$^{-1}$ for out of plane graphite foil and $\lambda \sim 2.3$ W. m$^{-1}$. K$^{-1}$ for silicon) may be responsible for a change in the growth mechanism of MgH$_2$. In particular, the role of heat dissipation at the film surface is expected to be a crucial effect in such mechanisms. We observed that, after the first desorption-absorption hydrogen cycle, the texture has disappeared due to recrystallization. At the macroscopic level, the rolling of the polyimide film occurring when it is taken out from the substrate holder in the deposition chamber suggests that the as-deposited $MgH_2$ film is in tensile stress. With currently available data, we cannot deduce or rule out the occurrence of a similar mechanical stress for the film deposited on graphite. In fact, no folding or rolling occurs after film deposition on graphite but this can also be explained by the large difference in substrate thicknesses (400 µm for graphite and 12.5 µm for polyimide). While the as-deposited films are strongly adherent to both substrates, the two types of films evolve in a very different fashion upon cycling. On the one hand, the graphite-supported film remains bonded after several hydrogen desorption-absorption cycles. On the other hand, the polyimide-supported film shows delamination upon cycling. While we did not perform mechanical characterization, we can provide a qualitative explanation based on the observed difference in behavior – the film on graphite remains bonded after hydrogen cycling while the film on polyimide does not. The graphite sheet, which is obtained from the compression of expanded graphite, is a highly anisotropic porous material allowing a certain amount of sliding among layers – and, hence, mechanical stress dissipation. In contrast, polyimide has a tensile strength that is 2 orders of magnitude higher than that of graphite. As a result, the former is less able to adapt to volume changes occurring when the film transforms from hydride to metal (theoretical $\Delta V/V \approx -30\%$) [21]. The film-polyimide interface is hence subjected to high strains upon hydrogen cycling. Finally, as an important potential explanation, we note that graphite foils are highly chemically and mechanically stable. On the other hand, thermal curing of the polyimide with associated change in the viscoelastic properties, can result from a prolonged operation in the vicinity of the glass transition temperature (300°C – 400°C, depending on supplier/ synthesis route). The present paper reports the investigation of two films of similar thickness (~1.8 µm) but deposited on different substrates. The high adherence over cycling of the $MgH_2$ film deposited on graphite raises complementary questions that could be the subject of a more comprehensive investigation: in particular, the impact of the thickness ratio between the film and the substrate as well as microscopic characterization of the film/substrate interface as a function of the number of hydrogen cycles (including for even larger amount of cycles). With the goal to provide information on the film/substrate interface, TEM analysis was attempted but did not yield data of sufficient quality. This is due to difficulties in preparing the TEM grid as the samples under study are sensitive to air. Moreover, the hydride films were also found to degrade under the electron beam. Regarding the film stability, note that Dura *et al*. [30] and Mooij *et al.* [26] showed that the stability of Mg film could be explained by voids formation upon dehydrogenation.

## 5. Conclusion

Reactive plasma sputtering was used to deposit hydrogenated magnesium films on flexible polyimide foils and graphite sheets. While the as-deposited films strongly adhere to the substrates, the film deposited on polyimide showed delamination and pulverization upon hydrogen cycling. Further study could focus on modifying the polyimide surface and/or applying a buffer layer to ensure stable adherence. Regarding the film deposited on graphite, upon multiple hydrogen desorption/absorption cycles, it retains its full hydrogen capacity and good bonding to the substrate. Microstructural investigations of the cycled sample showed the coarsening of the film accompanied by a loss of preferential orientation. A possible development path would be to optimize the film to substrate thickness ratio in order to maximize the hydrogen gravimetric capacity. The PAPVD process offers the benefit of a weaker sensitivity to oxidation of the freshly deposited hydride film (in contrast to metallic films such as Mg which are very sensitive to such oxidation processes). Moreover, from an industrial perspective, a single-stage process such as the PAPVD strategy used here is often associated with cost and time efficiency. We believe that the present proof of concept of direct deposition of hydride films on low-density flexible substrates can be implemented on a large scale. While the hydrogen capacity of the films considered here is low for conventional storage applications, it should be noted that no optimization was performed in the present paper. Therefore, possible improvements can be foreseen provided the key driving parameters leading to efficient hydrogen storage are identified. Despite such possible optimization, it is unlikely that such flexible films compete with bulk materials for hydrogen storage. Yet, in any case, the use of flexible supporting materials opens perspectives for specific applicative conditions. In particular, in the context of research efforts to minimize the volume of storage and conversion devices, flexible materials could pave the way for such developments. Indeed, such flexible substrate, which can be rolled up to form low volume systems while ensuring reasonable hydrogen storage capacity can be envisioned. Similarly, such materials providing means to design storage devices to fit applications with strong constraints regarding shape and volume. In this context, the use of a single-stage technique opens up new possibilities for the use of hydrides in emerging applications.

**Acknowledgements**

This work was financially supported by the French National Agency for Research under the grant MARIA'S STORY ANR-12-JS09-0011 and the French Carnot program "Energies du Futur" (D9: ATHOS PROJECT). Authors thank Emmanuel Verloop, Sébastien Pairis and Olivier Leynaud from CNRS - Institut Néel for their technical help with volumetric measurements, SEM and XRD characterizations respectively.